\RequirePackage{lineno}
\documentclass[twocolumn,aps,prc,showpacs,superscriptaddress,preprintnumbers,floatfix,nofootinbib]{revtex4}
\usepackage{epsfig,graphics}
\usepackage{graphicx}
\usepackage{dcolumn}
\usepackage{bm}
\usepackage{amsmath}
\usepackage[usenames]{color}
\usepackage{ulem} 

\begin{document}

\title{Dynamical effects of spin-dependent interactions in low- and intermediate-energy heavy-ion reactions}
\author{Jun Xu}
\email{xujun@sinap.ac.cn} \affiliation{Shanghai Institute of Applied
Physics, Chinese Academy of Sciences, Shanghai 201800, China}
\affiliation{Kavli Institute for Theoretical Physics China, CAS,
Beijing 100190, China}
\author{Bao-An Li}
\affiliation{Department of Physics and Astronomy, Texas A$\&$M
University-Commerce, Commerce, TX 75429-3011, USA}
\author{Wen-Qing Shen}
\affiliation{Shanghai Institute of Applied Physics, Chinese Academy
of Sciences, Shanghai 201800, China}
\author{Yin Xia}
\affiliation{Shanghai Institute of Applied Physics, Chinese Academy
of Sciences, Shanghai 201800, China}
\affiliation{University of
Chinese Academy of Sciences, Beijing 100049, China}

\begin{abstract}
It is well known that non-central nuclear forces, such as the spin-orbital coupling and the tensor force,
play important roles in understanding many interesting features of nuclear structures.
However, their dynamical effects in nuclear reactions are poorly known since only the spin-averaged observables are
normally studied both experimentally and theoretically. Realizing that spin-sensitive observables in nuclear reactions may carry useful information about the
in-medium properties of non-central nuclear interactions, besides earlier studies using the time-dependent Hartree-Fock approach to understand effects of spin-orbital coupling on
the threshold energy and spin polarization in fusion reactions, some efforts have been made recently to
explore dynamical effects of non-central nuclear forces in intermediate-energy heavy-ion collisions using transport models.
The focuses of these studies have been on investigating signatures of the density and isospin dependence of the
form factor in the spin-dependent single-nucleon potential. Interestingly, some useful probes were identified in the model studies while so far there is still no data to compare with.
In this brief review, we summarize the main physics motivations as well as the recent progress in understanding the spin dynamics and identifying spin-sensitive observables in
heavy-ion reactions at intermediate energies. We hope the interesting, important, and new physics potentials identified
in the spin dynamics of heavy-ion collisions will stimulate more experimental work in this direction.
\end{abstract}

\pacs{25.70.-z, 
      21.30.Fe, 
      21.10.Hw  
      }

\maketitle

\section{Introduction}
\label{introduction}

Understanding novel features of the fundamental nuclear forces and properties of
strongly interacting matter under extreme conditions of density, temperature, spin, and isospin are
among the main goals of nuclear physics. Heavy-ion collision (HIC) experiments play an
important role in achieving these goals. Indeed, great achievements have been made using
HICs at various beam energies from the sub-Colomb barrier to the highest energy available at
the large hadron collider. In particular, terrestrial experiments using intermediate-energy HICs have led to strong constraints on the equations
of state of hadronic matter~\cite{Dan02} and neutron-rich nucleonic matter~\cite{Bar05,Li08}.

Theoretical studies have shown recently that some spin-sensitive observables of HICs can be used to
explore the in-medium properties of non-central nuclear forces. The spin-dependent nuclear interactions are important for
explaining several interesting features of nuclear structure~\cite{Lia15}, such as the varying
magic numbers and the shell evolution with the isospin asymmetry of finite nuclei. However, the strength,
density, and isospin dependence of the nuclear spin-orbit
coupling are still uncertain (see Sec.~\ref{so}). Moreover, the
tensor force can modify the magic number of nuclei and is an
important source of the nucleon-nucleon short-range correlation. The latter is
related to many interesting phenomena in nuclear physics (see
Sec.~\ref{Ntensor}). More studies on in-medium properties of the spin-orbit coupling and tensor force
are thus vey much needed. HICs provide flexible ways of adjusting the conditions of the nuclear medium and may also lead
to new spin-dependent phenomena. For example, the so-called ``Spin
Hall Effect"~\cite{SHE1,SHE2,Hir99}, which affects the dynamics of
spin-up and spin-down particles differently as a result of the
spin-orbit coupling, is expected to be a general feature in any spin transport process.
It thus might be interesting to test if such phenomenon can also happen in HICs.

Considerable efforts using the time-dependent Hartree-Fock (TDHF) model, the spin- and isospin-dependent Boltzmann-Uehling-Uhlenbeck (SIBUU)
transport model, and the quantum molecular dynamics (QMD) model have
been devoted to exploring the spin dynamics in HICs (see
Sec.~\ref{reactions}). Indeed, some interesting phenomena were found. For example, it was found that the inclusion of the
spin-dependent nuclear interaction may affect the fusion threshold,
generate the spin twist during the collision process, and lead to
the spin splitting of nucleon collective flows (see Sec.~\ref{reactions}).
Future comparisons with relevant experimental data may help extract properties of the in-medium
spin-dependent nuclear force. Here we review briefly the main physics motivations and recent findings of
studying the spin-dependent dynamics and observables in low- and intermediate-energy
HICs. A major goal of this article is to stimulate more experimental work in this direction.

\section{Spin-related nuclear force}
\label{sforce}

Based on the one-boson-exchange picture~\cite{Machleidt}, nuclear
force can be understood by exchanging mesons between nucleons. Exchanging the scalar $\sigma$
meson and vector $\omega$ meson leads to respectively the attractive
and repulsive central nuclear force as well as the spin-orbit
interaction, while exchanging the $\pi$ meson and $\rho$ meson leads
to respectively the long-range and short-range nuclear tensor force.
Although in free space the bare nuclear force is well constrained by
the nucleon-nucleon scattering data, the in-medium nuclear
interactions, especially the nuclear spin-orbit interaction and
tensor force, are still quite uncertain.
The in-medium nuclear interactions can be studied by using microscopic many-body
theories or phenomenological models, such as the non-relativistic Skyrme-Hartree-Fock
(SHF) model and the relativistic mean-field (RMF) model. In the
following, we will discuss the effective spin-dependent nuclear
force based on the energy-density functional in the phenomenological
approach.

\subsection{Nuclear spin-orbit interaction}
\label{so}

The nuclear spin-orbit interaction was first introduced to explain
the magic numbers of nuclei~\cite{May48,Hax49}. Nuclei with numbers
of neutrons or protons equal to the magic numbers are more stable,
and this reflects the special shell structure of a nucleus. Although
even a simple harmonic potential leads to the shell structure of
nucleon energy levels inside nuclei, the spin-orbit coupling is
essential to reproduce the correct magic number.

In the Skyrme interaction, the effective spin-orbit force between
two nucleons at position $\vec{r}_1$ and $\vec{r}_2$ can be
expressed as~\cite{Vau72}
\begin{equation}\label{vso}
V_{so} = i W_0 (\vec{\sigma}_1+\vec{\sigma}_2) \cdot \vec{k} \times
\delta(\vec{r}_1-\vec{r}_2) \vec{k}^\prime.
\end{equation}
In the above, $W_0$ is the spin-orbit coupling constant,
$\vec{\sigma}_1$ and $\vec{\sigma}_2$ are the pauli matrices for the
two nucleons, $\vec{k}=-i(\nabla_1-\nabla_2)/2$ is the relative
momentum operator acting on the right side with $\nabla_1$ and
$\nabla_2$ acting on the first and second nucleon, respectively, and
$\vec{k}^\prime$ is its complex conjugate acting on the left. From
the conventional Hartree-Fock method, the spin-orbit single-particle potential can
be obtained based on the above effective spin-orbit force
\begin{equation}\label{uso_shf}
U_q^{so} = \vec{W}_q \cdot (-i\nabla \times \vec{\sigma}),
\end{equation}
where
\begin{equation}\label{ff_shf}
\vec{W}_q = \frac{W_0}{2}(\nabla \rho + \nabla \rho_q)
\end{equation}
is the form factor of the spin-orbit potential, with $q=\text{n}$ or
p being the isospin index and $\rho$ being the nucleon number
density. Taking the operator $-i\nabla$ as the momentum $\vec{p}$,
the right-hand side of Eq.~(\ref{uso_shf}) has the form of $(\vec{r}
\times \vec{p}) \cdot \vec{\sigma}$ with $\vec{W}_q$ playing the
role of $\vec{r}$, and this is why it is called the spin-orbit
potential. By solving the Schr\"odinger equation with the
single-nucleon Hamiltonian
\begin{equation}\label{sph}
h_q = -\nabla \cdot\left(\frac{1}{2m^\star_q}\nabla\right) + U_q +
U_q^{so}
\end{equation}
with $m^\star_q$ being the effective nucleon mass and $U_q$ being
the spin-independent potential, the single-nucleon spectrum in a
spherical closed-shell nucleus can be obtained.

In the RMF model, Dirac equation is solved where the spin of nucleon
is treated explicitly with nucleon wave functions for different spin
states~\cite{Rin96}. The studies of SHF and RMF models on nuclear structure were reviewed in Ref.~\cite{Ben03}, and here we compare the effective spin-orbit potentials from both the relativistic and non-relativistic
approach. With non-relativistic expansion of the Dirac equation, the
form factor of the nucleon effective spin-orbit potential in the RMF
model can be expressed in the form of~\cite{Rei89,Sul03}
\begin{equation}\label{ff_rmf}
\vec{W}_{RMF} = \frac{1}{(2m-C_{eff}\rho)^2} C_{eff} \nabla \rho,
\end{equation}
where $m$ is the nucleon mass and the coefficient $C_{eff}$ is
related to the coupling strength and mass of the scalar $\sigma$
meson and the vector $\omega$ meson, i.e.,
\begin{equation}
C_{eff} = \frac{g_\sigma^2}{m_\sigma^2} +
\frac{g_\omega^2}{m_\omega^2}.
\end{equation}

The form factors of the spin-orbit potential in the SHF model
(Eq.~(\ref{ff_shf})) and the RMF model (Eq.~(\ref{ff_rmf})) are
different. First of all, the spin-orbit coupling strength is a
constant in the SHF model, but the effective coupling strength
depends on the density in the RMF model. Implementing an additional
density-dependent effective nucleon-nucleon spin-orbit interaction
with a coupling constant $W_1$, the authors of Ref.~\cite{Pea94} got
addition contributions to the form factor as
\begin{eqnarray}
\vec{W}^\rho_{q}&=&\frac{W_1}{2} [c\rho \nabla(\rho-\rho_{q})
+(2+c)(2\rho_{q})^c\nabla\rho_{q}] \nonumber
\\ &+&\frac{W_1}{4}c\rho^{c-1} (\rho-\rho_{q})\nabla\rho,
\end{eqnarray}
with $c$ mimicking the density dependence. The above form was tested
in Ref.~\cite{Pea94} in a semi-infinite nuclear matter with
parameters fitted to the RMF interaction. It was found that the
general features of the RMF model were then reproduced with this
non-relativistic density-dependent spin-orbit interaction. Nevertheless, the density dependence of the spin-orbit coupling is still
largely unknown so far, and it is related to many interesting
phenomena in nuclear structure studies~\cite{Tod04,Gra09,Sor13}.
Second, the spin-orbit couplings from the SHF and RMF approach have
different isospin dependence, i.e., in the SHF approach the
spin-orbit coupling is stronger for nucleons of the same isospin,
while in the RMF approach the coupling strength is the same for
neutrons and protons. This feature impacts descriptions of
properties of neutron-rich nuclei, e.g., the kink in the evolution
of the charge radii for lead isotopes. It was shown that the weak
isospin dependence of the spin-orbit coupling in the RMF approach
can better explain the kink than the conventional SHF functional.
However, if the form factor in the latter approach was modified
to~\cite{Rei95,Sha95}
\begin{equation}
\vec{W}_{q} =\frac{W_0}{2}(1+\chi_{w})\nabla \rho_q +
\frac{W_0}{2}\nabla \rho_{q^\prime},~~(q \ne q^\prime)
\end{equation}
a similar kink can be reproduced with $\chi_w \approx
0.1$~\cite{Sha95}, corresponding to the case with very small Fock
contribution of the spin-orbit interaction. Similar efforts were
made by using a modified SHF functional to reproduce the isospin
dependence of the spin-orbit field in semi-infinite nuclear matter
with different neutron excesses~\cite{Ons97} and in neutron-rich
nuclei~\cite{Pea01} from a relativistic approach. In
Ref.~\cite{Lal98}, the isospin dependence of the spin-orbit coupling
was compared in light drip line nuclei from the relativistic mean
field theory and the non-relativistic Skyrme model. Furthermore, it was observed that the commonly used Skyrme functional
of the spin-orbit splitting overestimated the central density and
the spin-orbit splitting of neutron drops~\cite{Pud96}, calling for
new functionals of the spin-orbit coupling. The proton energy
splitting of $h_{11/2}$ and $g_{7/2}$ outside the $Z=50$ closed
shell increases with neutron excess, corresponding to the decreasing
strength of the nuclear spin-orbit interaction~\cite{Sch04}. The
studies so far seem to favor a weak isospin dependence of the spin-orbit
coupling. However, since the isospin dependence of the
spin-orbit coupling, which is important in nuclear surfaces, is often
coupled with its density dependence, it is still not well settled
yet.

Based on the above discussion, we proposed a general form of the
form factor of the spin-orbit coupling by taking both the density
and isospin dependence into account
\begin{equation}\label{Wgeneral}
\vec{W}_{q}
=\frac{W_0}{2}\left(\frac{\rho}{\rho_0}\right)^\gamma(a\nabla \rho_q
+ b\nabla \rho_{q^\prime})~~(q \ne q^\prime).
\end{equation}
The above form is artificially constructed and includes the main
physics for simplicity purpose. In the above form, $\gamma$ is used
to mimic the density dependence of the spin-orbit coupling while
fixing its strength at saturation density $\rho_0$ to be $W_0$. $a$
and $b$ are parameters to vary the isospin dependence of the
spin-orbit coupling, with $a=2$ and $b=1$ corresponding to the case
of the standard SHF approach and $a=b$ corresponding to the case of
the RMF approach. The values of $\gamma$, $a$, and $b$ are still
uncertain according to the above discussions. For the strength of
the spin-orbit coupling $W_0$, efforts have been made to extract its
information from ground-state properties of various nuclei. Recent
studies have shown that the spin-orbit coupling and the tensor
force, which will be discussed in the next subsection, should be
considered simultaneously to describe the spin-orbit splitting and
single-nucleon spectra of nuclei. Based on the Skyrme functional and
taking the uncertainties of the tensor force into account, the
strength of the spin-orbit coupling is approximately $80-150$ MeV
fm$^5$, from fitting the properties of light to heavy
nuclei~\cite{Les07,Zal08,Ben09}.

The single-nucleon Hamlitonian of Eq.~(\ref{sph}) is adequate to
describe the ground-state properties of spherical closed shell
nuclei. For open shell nuclei, one needs to consider an additional
spin-dependent potential using the spin-current density
$\vec{J}$ from Eq.~(\ref{vso})
\begin{equation}\label{uj}
U_q^J = -\frac{W_0}{2} \nabla \cdot (\vec{J} + \vec{J}_q).
\end{equation}
$\vec{J}$ is actually the vector component of the spin-current
density tensor $J_{\mu\nu}$, see, e.g., Ref.~\cite{Les07} for more detailed
discussions. For deformed nuclei, not only the time-even potentials
(Eqs.~(\ref{uso_shf}) and (\ref{uj})) but also the time-odd
potentials should be considered~\cite{Eng75}. Starting from the
effective nucleon-nucleon spin-orbit interaction (Eq.~(\ref{vso}))
and taking Eq.~(\ref{Wgeneral}) into consideration, the general form
of the time-even and time-odd spin-dependent potentials can be
written as
\begin{eqnarray}
U_q^{s-even} &=& -
\frac{W_0}{2}\left(\frac{\rho}{\rho_0}\right)^\gamma [\nabla \cdot
(a\vec{J}_q
+ b\vec{J}_{q^\prime}) ] \nonumber \\
&+& \frac{W_0}{2}\left(\frac{\rho}{\rho_0}\right)^\gamma(a\nabla
\rho_q + b\nabla \rho_{q^\prime})
\cdot (\vec{p} \times \vec{\sigma}) , \label{useveng}\\
U_q^{s-odd} &=& -
\frac{W_0}{2}\left(\frac{\rho}{\rho_0}\right)^\gamma\vec{p}
\cdot[\nabla
\times (a\vec{s}_q+ b\vec{s}_{q^\prime}) ] \nonumber\\
&-&\frac{W_0}{2}\left(\frac{\rho}{\rho_0}\right)^\gamma \vec{\sigma}
\cdot [\nabla \times (a\vec{j}_q + b\vec{j}_{q^\prime}) ], (q \ne
q^\prime) \nonumber \\\label{usoddg}
\end{eqnarray}
where $\vec{p}=-i\nabla$ is the momentum operator, and $\vec{s}$ and
$\vec{j}$ are spin density and current density, respectively. We
note that the time-odd potentials play an important role in the
dynamics of heavy-ion reactions, which will be discussed in
Sec.~\ref{reactions}.

\subsection{Nuclear tensor force}
\label{Ntensor}

The first strong evidence of the nuclear tensor force is from studying properties
of deuterons. In the non-relativistic
approach, the nuclear tensor force between two nucleons at position
$\vec{r}_1$ and $\vec{r}_2$ is often expressed with the tensor
operator written as
\begin{equation}
S_{12} = 3 \frac{(\vec{\sigma}_1 \cdot \vec{r})(\vec{\sigma}_2 \cdot
\vec{r})}{r^2}-(\vec{\sigma}_1 \cdot \vec{\sigma}_2),
\end{equation}
where $\vec{r}=\vec{r}_1-\vec{r}_2$ is the relative position vector.
One thus sees that whether the tensor force is attractive or
repulsive depends on the relative direction between the spin and the
relative position vector, i.e., $S_{12}>0$ for $\vec{\sigma}_{1(2)}$
parallel to $\vec{r}$ and $S_{12}<0$ for $\vec{\sigma}_{1(2)}$
perpendicular to $\vec{r}$.

The tensor term in the effective nuclear interaction was first
included in the Skyrme force~\cite{Sky58}, but afterwards it was
neglected due to its complex form. Recently it attracted renewed
interests. It has been found by Otsuka {\it et al.} that the nuclear
tensor force may affect the shell structure or even modify the magic
number of nuclei~\cite{Ots05,Ots06,Ots10}. The combined effects of
the spin-orbit coupling and the nuclear tensor force sometimes
hamper our understanding on both of them~\cite{Gau06,Col07}. Based
on the random phase approximation, effects of the tensor force on
the multipole response of magic nuclei have been studied, and a
large effect on the magnetic dipole states was
observed~\cite{Cao09}. Besides, it was proposed that the
spin-isospin excitation of finite nuclei may serve as a useful
observable to assess the strength of the tensor
force~\cite{Bai10,Bai11}. The existence of the tensor force may also
open a shell gap for large neutron numbers, having a consequent
implication for the synthesis of neutron-rich superheavy
elements~\cite{Suc10}.

Although the nuclear tensor force has no effect on the equation of
state of spin-saturated nuclear matter based on the studies at
mean-field level, it affects the properties of nuclear matter from
many-body calculation beyond the mean field. It was found that the
repulsive central force and the tensor force are two important
sources of nucleon-nucleon short-range correlation~\cite{Bet71}. The high-momentum tail of nucleon distribution
in nuclear matter as well as in finite nuclei was observed even at
zero temperature based on these
studies~\cite{Pan72,Fan84,Pei92,Cio96}. Great efforts have been made
in measuring the short-range nucleon-nucleon correlation and
extracting the ratio of nucleons in the high-momentum tail
experimentally~\cite{Tan03,CLAS06,Pia06,Sub08,Hen14} and
theoretically~\cite{Sch07,Alv08}, see, e.g., Ref.~\cite{Arr12} for a
review. In particular, it was found that the neutron-proton correlation
is much stronger than the correlation between neutron-neutron and proton-proton pairs~\cite{Pia06,Sub08,Hen14}, and this is mainly due to the
nuclear tensor force. The isospin dependence of the short-range
correlation can lead to interesting consequences, such as the
reduction of the kinetic contribution to the nuclear symmetry
energy~\cite{Vid11,Car12,Xu13} compared to the free Fermi gas
scenario. Since the symmetry energy at saturation density is constrained to be
around 30 MeV from many analyses, see, e.g., Ref.~\cite{Pom03}, the
isospin-dependent short-range correlation effectively increases the
potential contribution to the symmetry energy and thus the symmetry
potential effect, which may lead to enhanced isospin effects in
intermediate-energy HICs~\cite{Hen15,Li15}.

\section{Spin in nuclear reactions}
\label{reactions}

\begin{figure}[ht]
\includegraphics[scale=0.9]{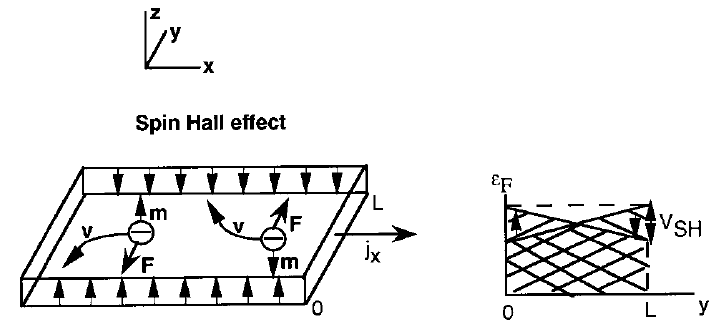}
\caption{Cartoon illustrating the spin hall effect. Taken from
Ref.~\cite{Hir99}. }\label{she}
\end{figure}

The spin hall effect was first predicted by M.I. Dyakonov and V.I.
Perel in 1971~\cite{SHE1,SHE2}, while the term "Spin Hall Effect"
was named by Hirsch in 1999~\cite{Hir99}. Considering the transport
of spin-up and spin-down particles with spin-orbit coupling, i.e.,
$U^{so}=-\vec{L} \cdot \vec{\sigma}$ with $\vec{L}$ being the
angular momentum and $\vec{\sigma}$ being the particle spin, the
spin-up (spin-down) particles tend to turn left (right) to couple
with the angular momentum and lower the energy, leading to the
splitting of final observables for different spin states, as shown
in Fig.~\ref{she}. In this section, we will discuss similar effects
in low- and intermediate-energy HICs based on the framework of TDHF,
BUU, and QMD, with more complicated forms of spin-orbit coupling
from spin-dependent nuclear interactions.

\subsection{TDHF model study}

The mean-field dynamics of nucleons in the TDHF model is described
by
\begin{equation}\label{TDHF}
i\frac{\partial}{\partial t} \phi_i = h \phi_i,
\end{equation}
where $\phi_i$ is the wave function of the $i$th nucleon and the
single-nucleon Hamlitonian is given by
\begin{equation}
h \phi_i = \frac{\delta E}{\delta \phi_i^\star},
\end{equation}
with $E$ being the energy functional of the nuclear system from
Hartree-Fock calculation. The single-nucleon Hamiltonian is
generally a function of nucleon number density $\rho$, spin density
$\vec{s}$, current density $\vec{j}$, spin-current density
$\vec{J}$, and so on, and their definitions in terms of the nucleon
wave function are
\begin{eqnarray}
\rho &=& \sum_i \phi^\star_i \phi_i,\\
\vec{s} &=& \sum_i \sum_{\sigma,\sigma^\prime} \phi^\star_i
\langle\sigma|\vec{\sigma}|\sigma^\prime\rangle \phi_i, \\
\vec{j} &=& \frac{1}{2i} \sum_i (\phi^\star_i \nabla\phi_i- \phi_i
\nabla\phi^\star_i),\\
\vec{J} &=& \frac{1}{2i} \sum_i \sum_{\sigma,\sigma^\prime}
(\phi^\star_i \nabla\phi_i- \phi_i \nabla\phi^\star_i)\times
\langle\sigma|\vec{\sigma}|\sigma^\prime\rangle,
\end{eqnarray}
with $\langle\sigma|\vec{\sigma}|\sigma^\prime\rangle$ being the
pauli matrix element. Numerically, these densities can be calculated
on the coordinate space grid and Eq.~(\ref{TDHF}) can be solved with
a fixed time step. For more details, we refer the reader to
Refs.~\cite{Hoo77,Dav81}. The TDHF framework works well for
low-energy heavy-ion reactions and in studing resonances dynamics.

\begin{figure}[ht]
\includegraphics[scale=0.9]{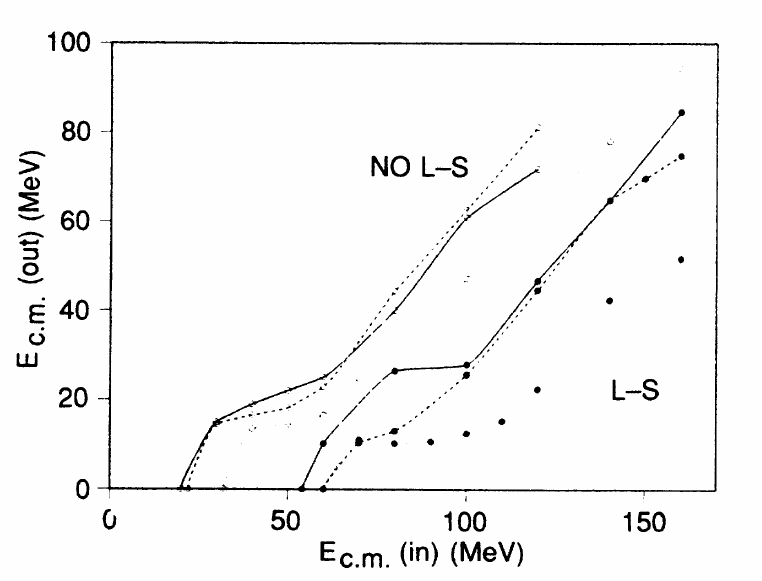}
\caption{ The relation between the out-going energy and incoming
energy in $\text{O}^{16}+\text{O}^{16}$ reaction from TDHF study.
The filled (open) circles are results with (without) spin-orbit
force, and the solid, dashed, and dotted curves represent results
from three different Skyrme forces. Taken from Ref.~\cite{Rei88}.
}\label{threshold}
\end{figure}

In the old calculations, the single-nucleon Hamiltonian was
generally calculated from the SHF model without the spin-orbit
interaction and time-odd terms. The spin-orbit force was first
introduced to the TDHF framework in Refs.~\cite{Uma86,Rei88,Uma89}.
It is interesting to see that the spin-orbit force enhances the
dissipation in the fusion reaction and transforms the relative
motions of the two nuclei into the internal excitations. The fusion
threshold energy in $\text{O}^{16}+\text{O}^{16}$ reaction is
increased by about a factor of $2$~\cite{Uma86,Rei88}, as shown in
Fig.~\ref{threshold} with three different parameterizations of
Skyrme force. The fusion cross section obtained from TDHF
calculation was increased after including the spin-orbit
force~\cite{Uma86}.

\begin{figure}[ht]
\includegraphics[scale=0.3]{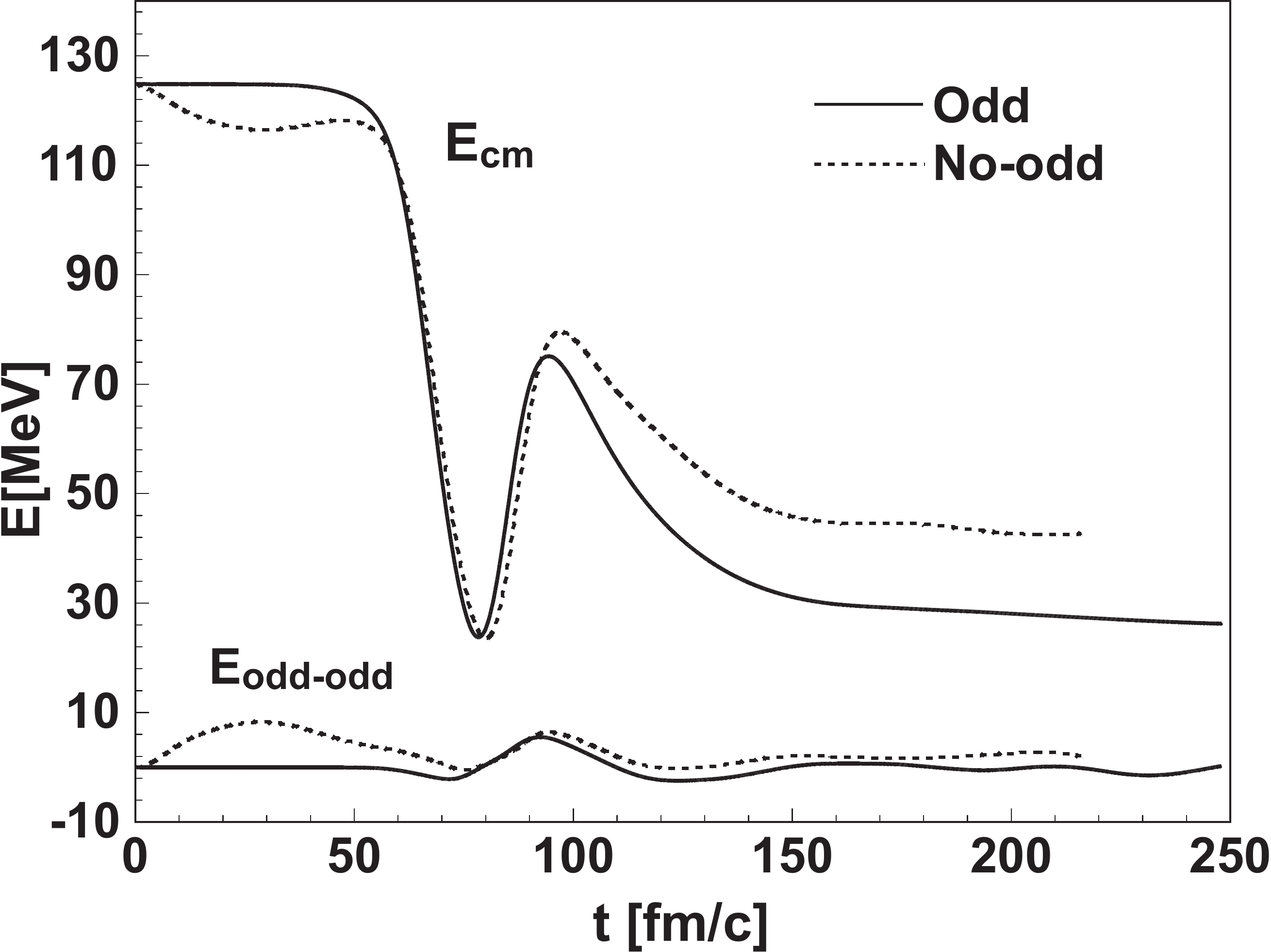}
\caption{Center-of-mass energy evolution in central
$\text{O}^{16}+\text{O}^{16}$ reaction with (solid lines) and
without (dashed lines) time-odd contributions from TDHF calculation.
The time window of the reaction process is from about 50 fm/c to 120
fm/c. Taken from Ref.~\cite{Mar06}. }\label{odd}
\end{figure}

\begin{figure}[ht]
\includegraphics[scale=0.3]{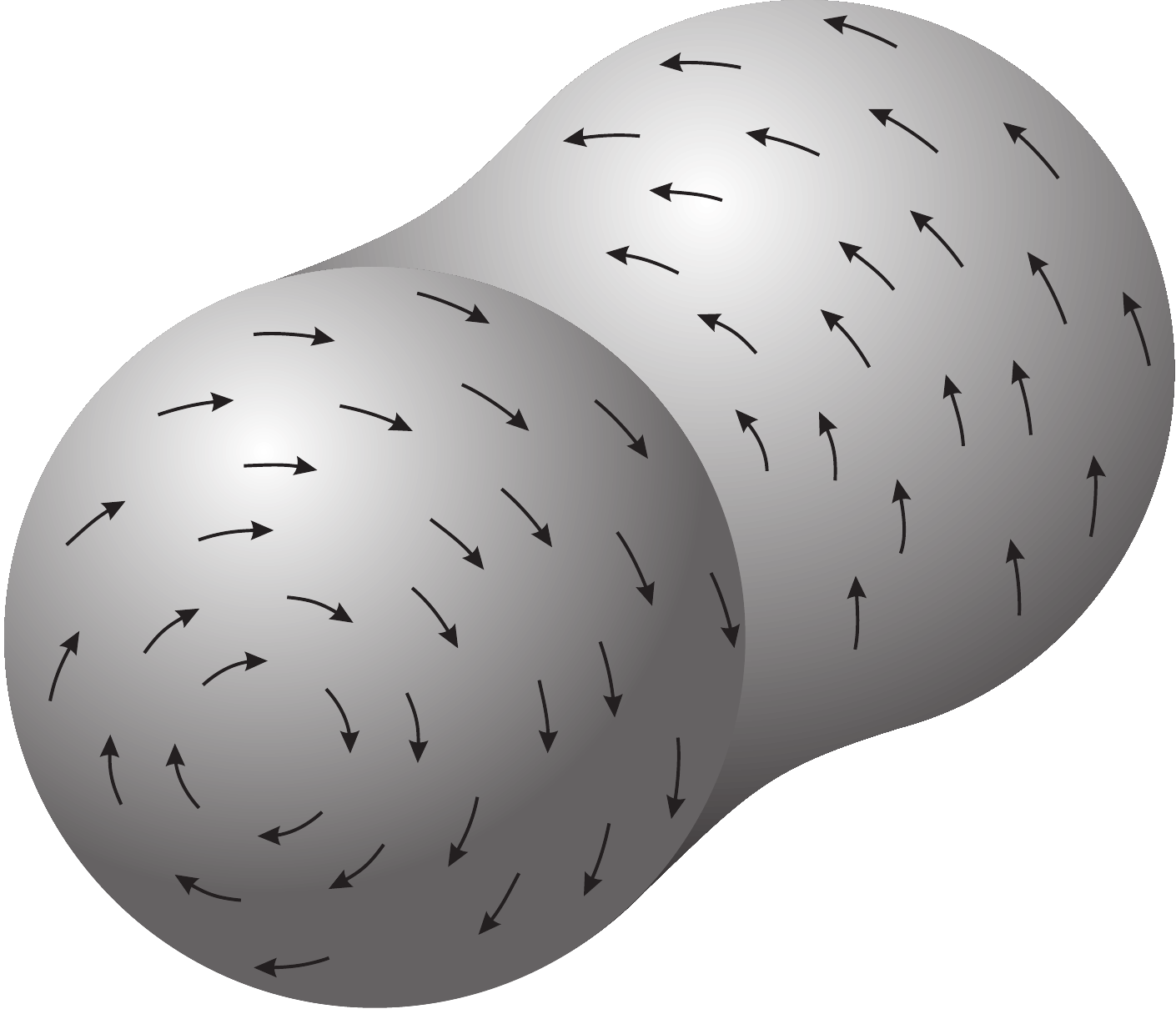}
\caption{Spin excitation in central $\text{O}^{16}+\text{O}^{16}$
reactions from TDHF study with both time-even and time-odd terms.
Taken from Ref.~\cite{Mar06}. }\label{twist}
\end{figure}

With only time-even contribution of the spin-orbit interaction,
i.e., the spin-orbit potential (Eq.~(\ref{uso_shf})) and the
potential with spin-current density $\vec{J}$ (Eq.~(\ref{uj})),
spurious spin twist can be generated in a free moving nucleus, as a
result of spin-orbit coupling. Obviously, this phenomena is not
reasonable as it depends on the reference frame. Considering that
all kinds of collision geometry can be realized in HICs, the
time-odd terms were further introduced in the TDHF calculation in
Refs.~\cite{Uma06,Mar06} to satisfy the invariance under Galilei
transformations. It is seen from Fig.~\ref{odd} that there is no
such spurious spin and the kinetic energy is a constant before 50
fm/c when the nuclei are moving freely, as a result of suppression
effect on the time-even terms from the time-odd terms. During the
reaction process, the real spin twist appears due to the
overwhelming effect of the time-odd terms on the time-even terms, as
shown in Fig.~\ref{twist}. At the end of the reaction, the energy of
outgoing nuclei is smaller with the time-odd terms as shown in
Fig.~\ref{odd}, indicating a stronger dissipation. Besides the spin
excitation, it was found that the fusion description was further
improved with the time-odd terms and the spin-current pseudotensor
contribution~\cite{Uma06}. A more detailed study on this topic was
done recently~\cite{Dai14a}, where it was found that the dissipation
is dominated by the time-even contribution of the spin-orbit force
at lower energies but by the time-odd terms at higher energies.

\begin{figure}[ht]
\includegraphics[scale=1.0]{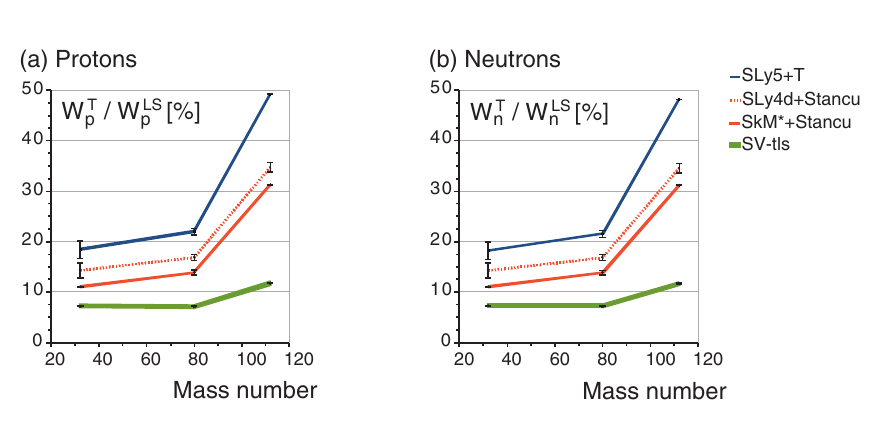}
\caption{(Color online) Ratios of the spin mean field from the
spin-current density representing the tensor force to that from the
spin-orbit force for protons (left) and neutrons (right) as a
function of the mass of reaction nuclei. Different parameterizations
of the Skyrme force are used in the study. Taken from
Ref.~\cite{Iwa11}. }\label{tensor}
\end{figure}

Besides the spin-orbit force, the additional contribution of the
spin-current density $\vec{J}$ was introduced in the TDHF
calculation representing the contribution from the tensor force in
Ref.~\cite{Iwa11}. It was found that the dissipation effect from the
tensor force is small compared with that from the spin-orbit
force~\cite{Iwa11,Dai14b}. However, the spin mean field can be
enhanced with the nuclear tensor force, and the enhancement becomes
important with the increasing mass of the colliding nuclei, as shown in
Fig.~\ref{tensor}, depending on the parameterization of the Skyrme
force. A more complete study by including the full Skyrme functional as well as the tensor force in the TDHF calculation was done very recently in Ref.~\cite{Ste15}. It was found that the Skyrme tensor force has non-negligible effects on low-energy heavy-ion dynamics and the fusion threshold energy.

\subsection{BUU model study}

The TDHF model works well in low-energy HICs, while the particle
emission and nucleon-nucleon scattering are still lacking. To
describe these effects in intermediate-energy HICs, BUU models and
QMD models are suitable candidates. In the BUU framework, the
Boltzmann equation is solved with test particle
method~\cite{Ber88,Won82}. In the previous studies, an
isospin-dependent BUU (IBUU) transport model has been used to
describe the isospin dynamics in intermediate-energy
HICs~\cite{Li08}. Recently, the spin degree of freedom of nucleons
and the spin-orbit interaction were incorporated in the IBUU model,
and the new model is dubbed as the spin- and isospin-dependent BUU
(SIBUU) model~\cite{SIBUU13,SIBUU14,SIBUU15,Xu14NPR,Xu14NT}.
In this section, we summarize the main results published originally in
Refs.~\cite{SIBUU13,SIBUU14,SIBUU15,Xu14NPR,Xu14NT}.

In the SIBUU model, each nucleon is assigned randomly a unit vector
representing the expectation value of its spin. In this way, the
spin projection of each nucleon at arbitrary direction can be easily
calculated. In the transport simulation, $z$ direction is set as the
beam momentum and $x$ direction is for the impact parameter. Since
the total angular momentum in non-central HICs is in the $y$
direction perpendicular to the reaction plane, i.e., $x-o-z$ plane,
it is reasonable to study the spin polarization in $y$ direction. We
thus determine the nucleons with spin projection on $+y$ ($-y$)
direction as the spin-up (spin-down) nucleons.

Considering the general form of the time-even and time-odd
spin-dependent potentials in Eqs.~(\ref{useveng}) and
(\ref{usoddg}), the time evolutions of the coordinate, momentum, and
spin degree of freedom are described by
\begin{eqnarray}
\frac{d\vec{r}}{dt} &=& \frac{\vec{p}}{m} +
\frac{W_0}{2}\left(\frac{\rho}{\rho_0}\right)^\gamma \vec{\sigma}
\times (a\nabla \rho_q +
b\nabla \rho_{q^\prime}) \nonumber \\
&-& \frac{W_0}{2}\left(\frac{\rho}{\rho_0}\right)^\gamma \nabla \times (a\vec{s}_q + b\vec{s}_{q^\prime}), \label{rt}\\
\frac{d\vec{p}}{dt} &=& - \nabla U_q - \nabla U_q^{s-even} - \nabla
U_q^{s-odd}, \label{pt}
\\
\frac{d\vec{\sigma}}{dt}
&=&W_0\left(\frac{\rho}{\rho_0}\right)^\gamma[(a\nabla \rho_q +
b\nabla \rho_{q^\prime}) \times \vec{p}] \times
\vec{\sigma} \nonumber \\
&-&W_0\left(\frac{\rho}{\rho_0}\right)^\gamma[\nabla \times
(a\vec{j}_q + b\vec{j}_{q^\prime})] \times \vec{\sigma}.
\label{sigmat}
\end{eqnarray}
One sees that the three degrees of freedom couple with each other.
The number density $\rho$, the spin density $\vec{s}$, the current
density $\vec{j}$, and the spin-current density $\vec{J}$ are
calculated from test particle method~\cite{Won82,Ber88,SIBUU14}.
Since the mixing of the long-range Fock contribution and the spin
interaction is a complex problem, the momentum dependence is not
included in the spin-independent potential $U_q$ for the moment. In
addition, the spin of nucleons are randomized after nucleon-nucleon
scatterings, by approximately taking the spin flip effect into
consideration~\cite{Ohl72,Lov81}.

\begin{figure}[ht]
\includegraphics[scale=0.45]{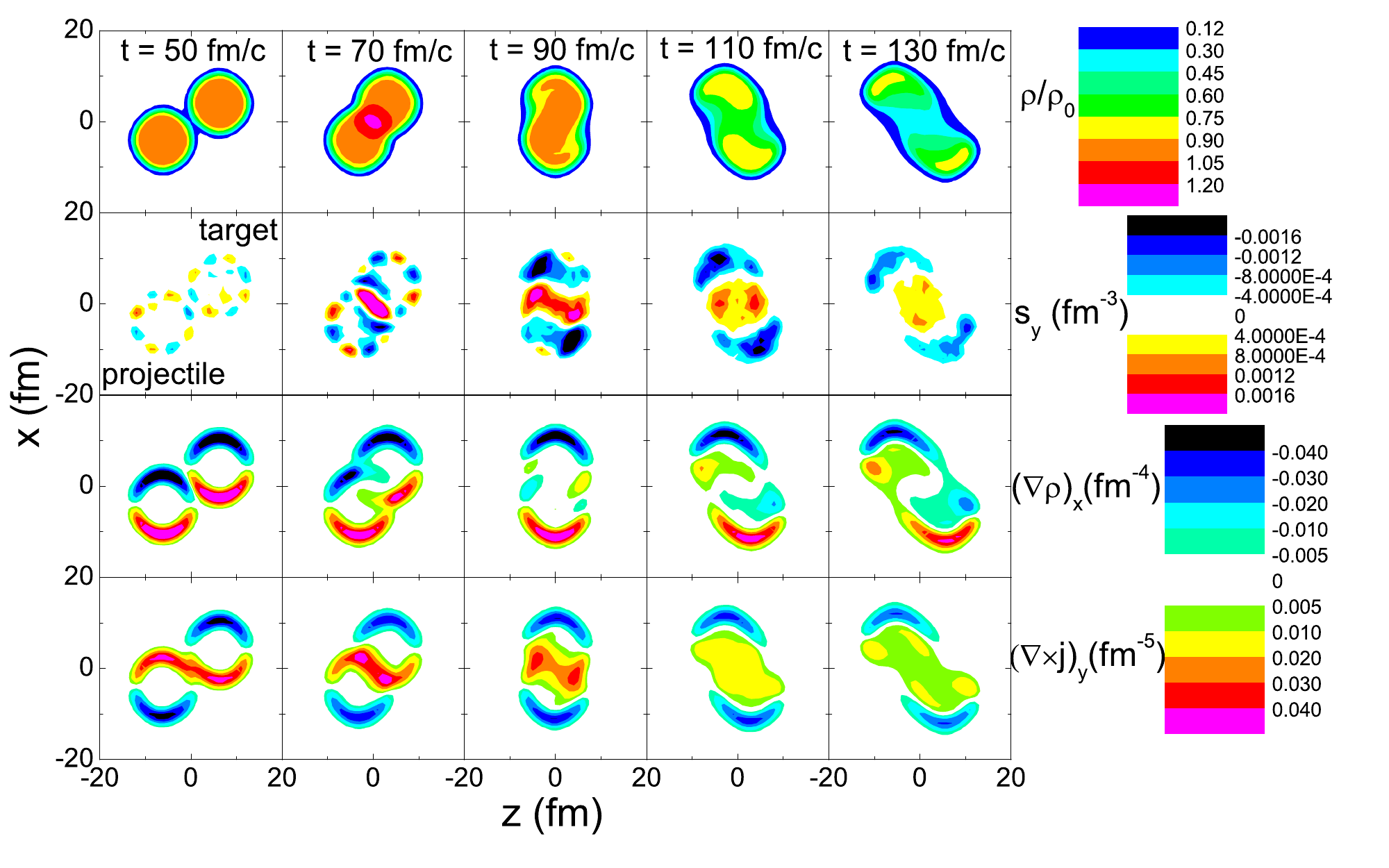}
\caption{(Color online) Time evolution of contours of reduced number
density $\rho/\rho_0$, $y$ component of the spin density $s_y$, $x$
component of the number density gradient $(\nabla \rho)_x$, and $y$
component of the curl of the current density $(\nabla \times
\vec{j})_y$ in non-central Au+Au collision at the beam energy of $50$
MeV. Taken from Ref.~\cite{SIBUU13}. }\label{den}
\end{figure}

The time evolution of relevant density contours from SIBUU
calculation are displayed in Fig.~\ref{den}. The gradient of number
density $\nabla \rho$ and the curl of the current density $\vec{j}$
show the strength of the time-even and time-odd spin-dependent
potential, respectively, and both of them are closely related to the
evolution of the number density shown in the first row of
Fig.~\ref{den}. The nucleon spin tends to be parallel to $\vec{p}
\times \nabla \rho$ from the time-even potential
(Eq.~(\ref{useveng})), while it tends to be parallel to $\nabla
\times \vec{j}$ from the time-odd potential (Eq.~(\ref{usoddg})).
The contributions from the time-even and time-odd potentials are
opposite to each other. One sees that before the two nuclei touch
each other there is no spin polarization as a result of the
cancellation of the time-even and time-odd potentials, consistent
with the foundings from TDHF studies. During the collision process,
the participant is polarized in the $+y$ direction, i.e., in the
direction of the total angular momentum, following the preference
direction of the time-odd potential. It is seen that the direction
of the spin polarization is consistent with that in Fig.~\ref{twist}
from TDHF calculation with both time-even and time-odd potentials.

\begin{figure}[ht]
\includegraphics[scale=0.8]{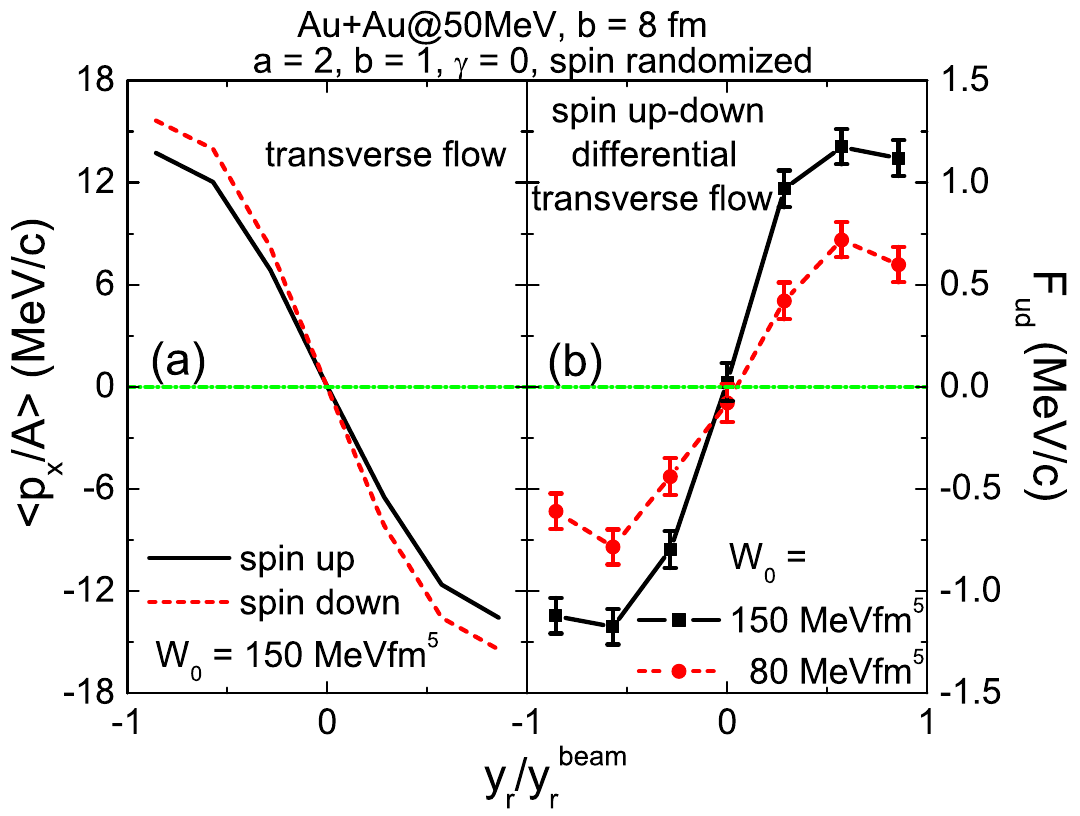}
\caption{(Color online) Transverse flow of spin-up and spin-down
nucleons (left) and spin up-down differential transverse flow
(right) with different strength of the spin-orbit coupling in
non-central Au+Au collisions at the beam energy of $50$ MeV. Taken
from Ref.~\cite{SIBUU13}. }\label{Fud}
\end{figure}

Transverse flow is one of the most important observables for
extracting the equation of state of produced matter and studying the
nuclear interaction in HICs~\cite{Dan85,Ber88,Dan02}. The left panel
of Fig.~\ref{Fud} displays the transverse flow of spin-up and
spin-down nucleons as a function of reduced rapidity
$y_r/y_r^{beam}$. We note that the target (projectile) nucleus is in
the $+x$ ($-x$) direction in Fig.~\ref{den}, which is different from
the conventional initialization, leading to the negative slope of
the transverse flow. However, this doesn't prevent the reader from
seeing the obvious splitting of transverse flow between spin-up and
spin-down nucleons. With a detailed orientation analysis, one can
find that again the time-odd potential dominates the effect, giving
the spin-up (spin-down) nucleons an attractive (repulsive)
potential. This can be understood in a naive picture that the
spin-up (spin-down) nucleons parallel (antiparallel) to the
direction of total angular momentum and thus feel an attractive
(repulsive) potential. One can further define the spin up-down
differential transverse flow as follows
\begin{equation}
F_{ud}(y_r) = \frac{1}{N(y_r)} \sum_{i=1}^{N(y_r)} \sigma_i (p_x)_i,
\end{equation}
where $\sigma_i$ is $1$ for spin-up nucleons and $-1$ for spin-down
nucleons, and $N(y_r)$ is the number of nucleons at rapidity $y_r$.
The above spin up-down differential transverse flow largely cancels
the effect from the spin-independent nuclear interaction while
preserves the information of the spin-dependent potential. Indeed,
the slope of $F_{ud}$ increases with increasing spin-orbit coupling
constant, indicating that it is a good probe of nuclear
spin-dependent interaction.

\begin{figure}[ht]
\includegraphics[scale=0.8]{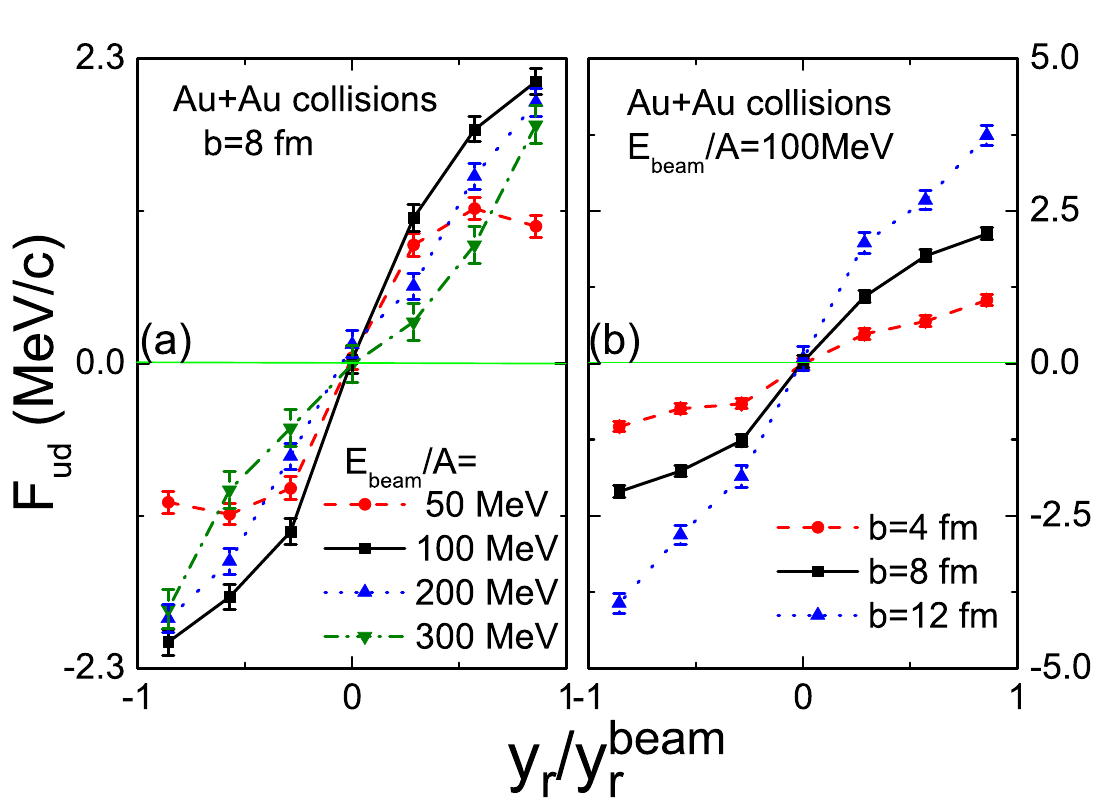}
\caption{(Color online) Dependence of the spin up-down differential
transverse flow on the beam energy (left) and the impact parameter
(right) in non-central Au+Au collisions. Taken from
Ref.~\cite{SIBUU14}.} \label{Fud_energy}
\end{figure}

The spin up-down differential transverse flow was further analyzed
in detail in Ref.~\cite{SIBUU14}. Figure~\ref{Fud_energy} displays
the dependence of $F_{ud}$ on the beam energy and the centrality. At
higher beam energies, the angular momentum is larger while the
nucleon-nucleon scattering is more violate, with the former
enhancing the spin-dependent potential while the latter washing out
part of the information of spin dynamics. The competition leads to a
maximum slope of $F_{ud}$ at the beam energy of about $100$ MeV, as
shown in the left panel of Fig.~\ref{Fud_energy}. Since the
spin-dependent potential is related to the density gradient and is
thus a surface effect, the slope of $F_{ud}$ increases with the
increasing value of the impact parameter, as shown in the right
panel of Fig.~\ref{Fud_energy}.

\begin{figure}[ht]
\includegraphics[scale=1.2]{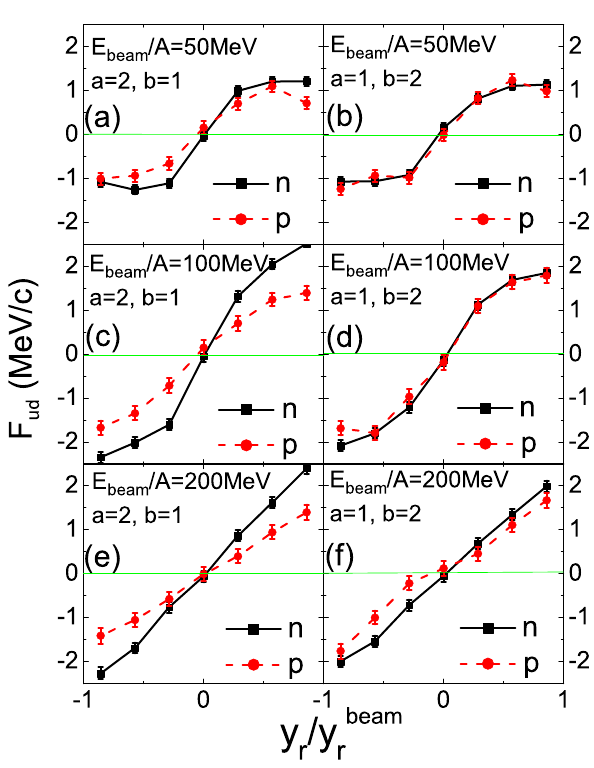}
\caption{(Color online) Spin up-down differential transverse flow of
neutrons and protons at different beam energies and with two typical
isospin dependence of the spin-orbit coupling. Taken from
Ref.~\cite{SIBUU14}.}\label{Fud_isospin}
\end{figure}

In the neutron-rich collision system where the relevant neutron
densities are larger than proton densities, the difference of the
spin up-down differential transverse flow of neutrons and protons can
be useful to probe the isospin-dependence of the spin-orbit coupling
in HICs. The analysis was carried out with a stronger isospin-like
coupling ($a=2$, $b=1$) and a stronger isospin-unlike coupling
($a=1$, $b=2$), and the resulting $F_{ud}$ were calculated at
different beam energies shown in Fig.~\ref{Fud_isospin}. A stronger
isospin-like spin-orbit coupling, which is exactly the case of SHF
interaction, leads to a larger $F_{ud}$ for neutrons than for
protons, while a stronger isospin-unlike coupling gives opposite
predictions or similar $F_{ud}$ for neutrons and protons. The effect
is appreciable from beam energy 50 MeV to 200 MeV, while the beam
energy of $100$ MeV is the optimized one due to the largest
magnitude of $F_{ud}$.

\begin{figure}[ht]
\includegraphics[scale=0.8]{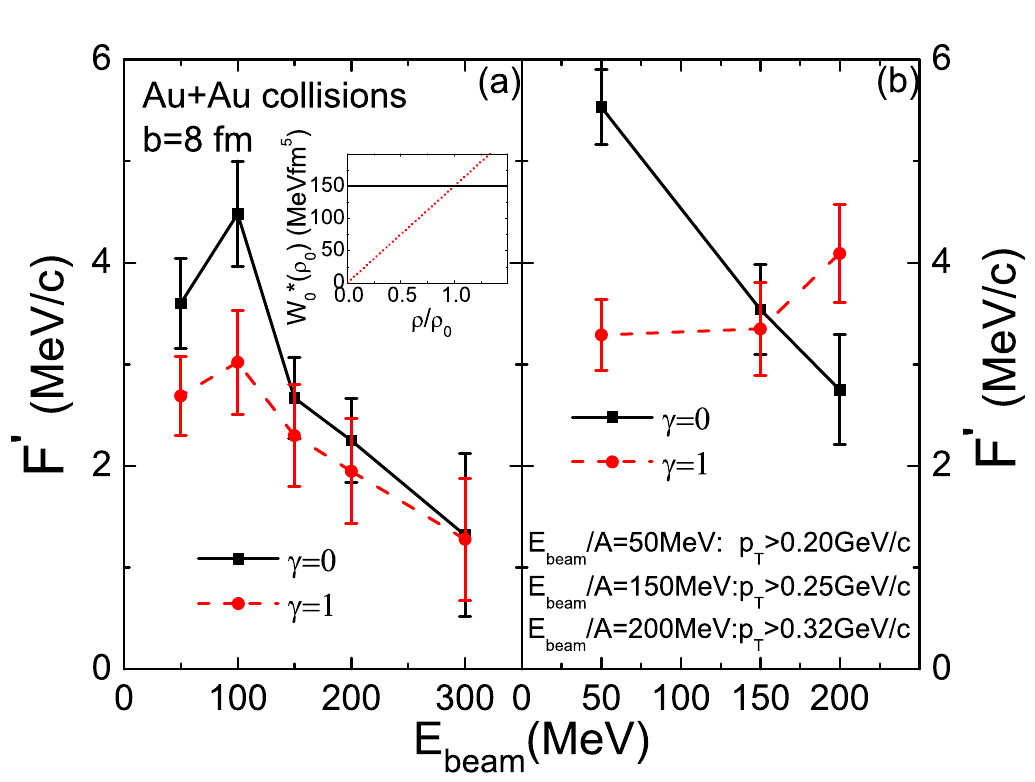}
\caption{(Color online) Slope parameter of the spin up-down
differential transverse flow $F^\prime$ without (left) and with
(right) high transverse momentum cut from different density
dependence of the spin-orbit coupling at different beam energies.
Taken from Ref.~\cite{SIBUU14}.}\label{Fud_density}
\end{figure}

The density dependence of the spin-orbit coupling has bothered many
nuclear physicists and hampered the understanding of nuclear
spin-orbit interaction in nuclear structure studies. Since HICs have
the advantage of constructing the system with designed density,
isospin, and momentum current, it might be helpful in extracting
useful information of the density dependence of the spin-orbit
coupling. As is known, nucleons of high transverse momentum ($p_T$)
emit early from the high-density phase in HICs, and the density of
the high-density phase increases with increasing beam energy. This
feature can be used to extract the density dependence of the
spin-orbit coupling, as illustrated in Fig.~\ref{Fud_density}.
Without high-$p_T$ cut, the slope of $F_{ud}$ can hardly be
distinguished as shown in the left panel of Fig.~\ref{Fud_density},
because nucleon emission from low-density phase, which is similar at
different beam energies, dominates the results. With high-$p_T$ cut,
the slope of $F_{ud}$ is smaller at lower collision energies but
larger at higher collision energies from a linearly increasing
spin-orbit coupling strength, compared to the case with a constant
one. In this way, the strength and the density dependence of the
spin-orbit coupling can be disentangled.

\begin{figure}[ht]
\includegraphics[scale=2.0]{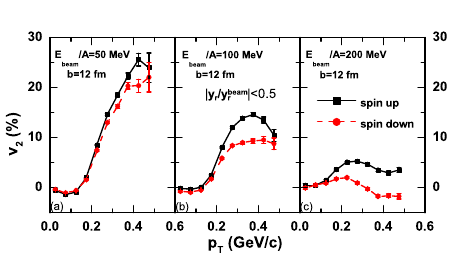}
\caption{(Color online) Transverse momentum dependence of the
elliptic flow of mid-rapidity nucleons in peripheral Au+Au
collisions at different beam energies. Taken from
Ref.~\cite{SIBUU15}.}\label{v2_pt}
\end{figure}

In non-central HICs, the azimuthal distribution of emitted nucleons
can always be expressed as
\begin{eqnarray}
E\frac{d^3N}{dp^3} &=& \frac{d^2N}{2\pi p_T dp_T dy_r}
[ 1 + 2v_{1}(y_r,p_T)\cos(\phi) \notag\\
&+& 2v_{2}(y_r,p_T)\cos(2\phi) +... ]
\end{eqnarray}
with $\phi=\tan^{-1}(p_y/p_x)$ being the azimuthal angle, and
$v_1=\langle\cos(\phi)\rangle$ and $v_2=\langle\cos(2\phi)\rangle$
are called the directed flow and elliptic flow, respectively. The
directed flow is similar to the transverse flow but it depends on
the flow angle rather than magnitude. The elliptic flow is positive
at lower energies, negative at intermediate energies, and becomes
positive again at higher energies. The positive elliptic flow means
more particles move in-plane than out-of-plane as a results of
hydrodynamics, while the negative elliptic flow is a result of the
squeeze-out effect on the expansion of participant matter by the
spectator nucleons~\cite{Dan02}. Despite of the complicated
dynamics, the elliptic flow serves as a useful probe of the
properties of nuclear matter formed in HICs and the nuclear
interaction. The transverse momentum dependence of $v_2$ of spin-up
and spin-down nucleons at mid-rapidity is displayed in
Fig.~\ref{v2_pt}. Except for the different behaviors of $v_2$ at
different beam energies, the large elliptic flow of spin-up nucleons
than spin-down nucleons is observed, especially at higher transverse
momentum as a result of the stronger spin-orbit coupling for
energetic nucleons. At the energy range considered, a more
attractive mean-field potential leads to a larger $v_2$ in
peripheral HICs, consistent with the effect of spin-dependent
potential on the spin splitting of transverse flow discussed above.

The above observables are for free nucleons. Experimentally it is
easier to detect charged particles rather than neutrons, leading to
difficulties of measuring the spin splitting of transverse flows for
protons and neutrons and identifying the isospin dependence of the
spin-orbit coupling. Of course the spin measurement is another
challenge which will be discussed in the next section. Once the
corresponding detectors are set up, the spin splitting of
observables for charged light clusters may be more easily measured.
For transport models with point-like particles, the dynamical
coalescence approach has been shown to be successful in studying the
hadronization in relativistic HICs~\cite{Gre03a,Gre03b} and light
cluster formation in intermediate-energy HICs~\cite{Che03a,Che03b}.
In this approach, the probability for nucleons to form a light
cluster is proportional to the nucleon Wigner function of the light
cluster~\cite{Che03a,Che03b}, and the proportional constant is the
statistical factor determined by the spin-isospin degeneracy. For
example, with explicitly knowing the isospin of nucleons, the
statistical factor for a neutron and a proton to form a deuteron is
$3/8$, while that for one neutron and two protons to form a $^3$He is
$1/12$. Since now the spin of each nucleon is also explicitly known,
the dynamical coalescence can be further improved by considering the
antisymmetrization of the product of spin and isospin wave
function. For example, the statistical factor for a spin-up neutron
and a spin-up proton to form a spin-up deuteron is $1/2$, while for
a spin-up neutron, a spin-up proton, and a spin-down proton to form
a spin-up $^3$He is $1/2$. This improvement has been applied to
study spin splitting observables for deuterons, tritons, and
$^3$He~\cite{SIBUU16}. It has been checked that after spin average
the results reproduce those without explicit spin treatment.

\begin{figure}[ht]
\includegraphics[scale=1.0]{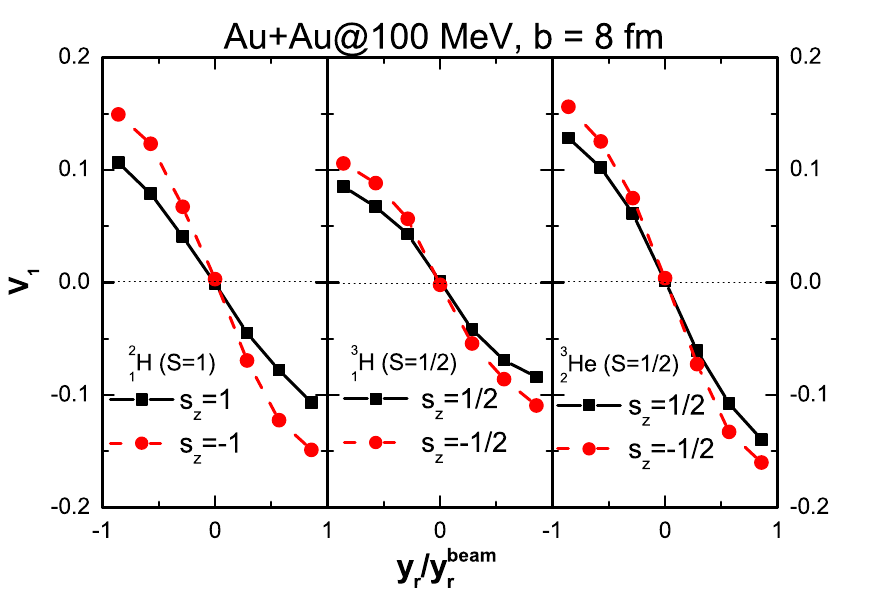}
\caption{(Color online) Directed flow of deuteron, triton, and
$^3$He of different spin states in non-central Au+Au collisions at
the beam energy of $100$ MeV. $s_z$ represents the spin state projecting on the $y$ direction perpendicular to the reaction plane. Taken from Ref.~\cite{SIBUU16}.
}\label{v1_cluster}
\end{figure}

Figure~\ref{v1_cluster} displays the spin splitting of the directed
flows for deuterons, tritons, and $^3$He in non-central Au+Au
collisions at the beam energy of $100$ MeV. The directed flow of
spin-down clusters is larger than that of spin-up ones. The spin
splitting of the directed flow is largest for deuterons due to its
large spin quantum number, i.e., $S=1$. The spin splitting
observables of tritons and $^3$He might be more easily measurable
for extracting the isospin dependence of the spin-orbit coupling
experimentally.

\begin{figure}[ht]
\includegraphics[scale=0.8]{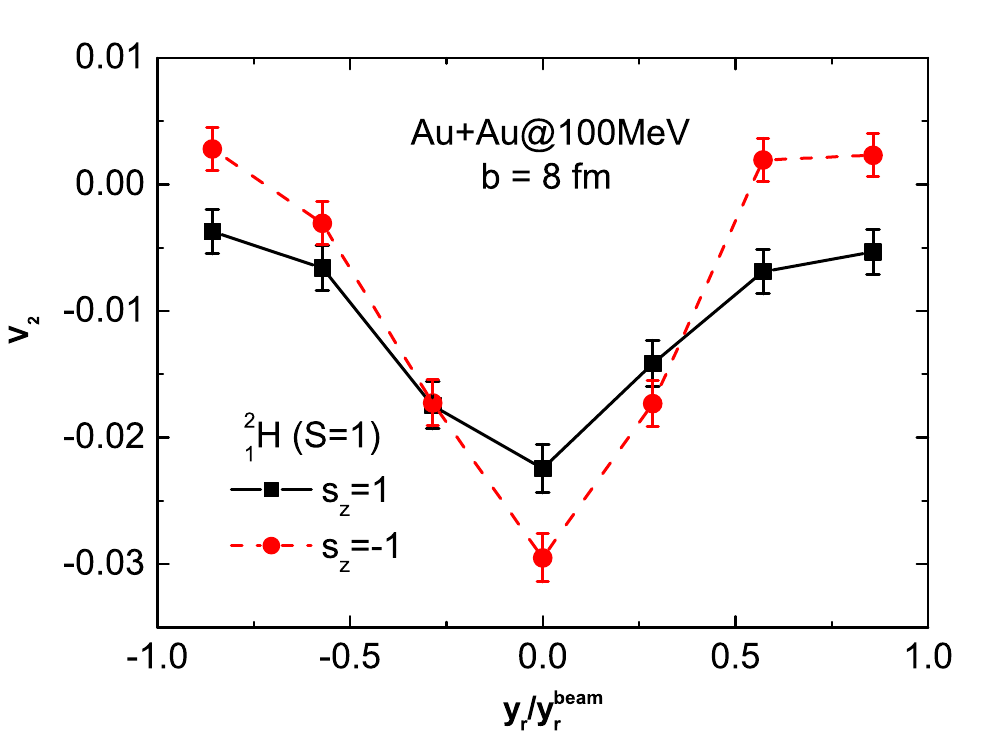}
\caption{(Color online) Elliptic flow of of spin-up and spin-down
deuterons in non-central Au+Au collisions at the beam energy of
$100$ MeV. $s_z$ represents the spin state projecting on the $y$ direction perpendicular to the reaction plane. Taken from Ref.~\cite{SIBUU16}.}\label{v2_cluster}
\end{figure}

The elliptic flow of different spin states of deuterons has been
illustrated in Fig.~\ref{v2_cluster}, in non-central Au+Au
collisions at the beam energy of $100$ MeV. It is seen that the
elliptic flow of spin-down deuterons is more negative at
mid-rapidity, but is slightly positive at large rapidity, indicating
an obvious spin splitting even taking the statistical error into
account. Again, the magnitude of the $v_2$ as well as its spin
splitting for deuterons is larger than that of free nucleons
according to Ref.~\cite{SIBUU15}, and might serve as a better
spin-dependent observable.

\begin{figure}[ht]
\includegraphics[scale=0.8]{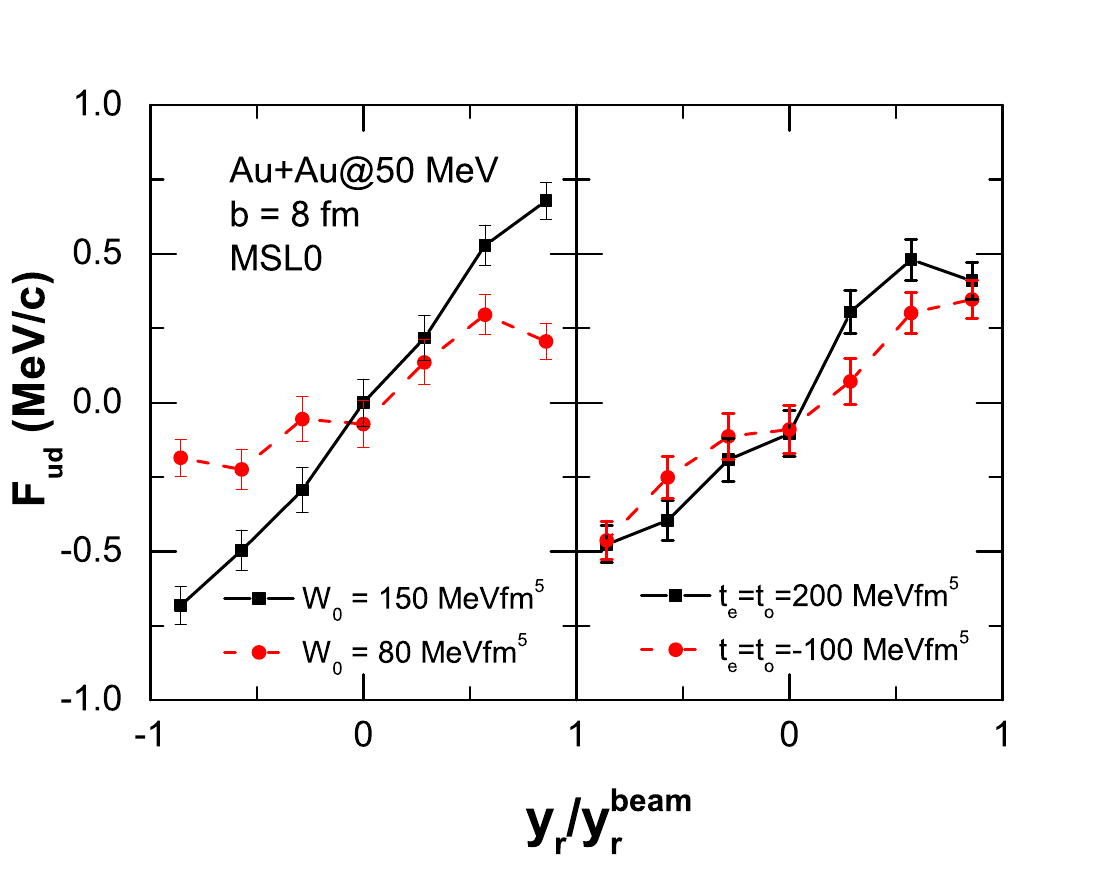}
\caption{(Color online) Spin up-down differential transverse flow
from full Skyrme calculation using MSL0 force without (left) and
with tensor force (right). Taken from Ref.~\cite{SIBUU16}.
}\label{full_tensor}
\end{figure}

Further preliminary calculation with full Skyrme functional has been
done. A standard Skyrme functional with MSL0
parameterization~\cite{MSL0} has been used in the calculation. The
detailed derivation and expression of full Skyrme functional with
both time-even and time-odd terms can be found in
Refs.~\cite{Eng75,Les07,Ben09}. The resulting spin up-down
differential transverse flow is shown in the left panel of
Fig.~\ref{full_tensor}. One can see the similar sensitivity of
$F_{ud}$ to the spin-orbit coupling strength although the magnitude
is a little smaller, compared to the result shown in Fig.~\ref{Fud}
where only the spin-orbit coupling is applied. To investigate the
effect of nuclear tensor force on the spin dynamics of
intermediate-energy HICs, a zero-range tensor force of the form
\begin{eqnarray}
v_t(\vec{r}) &=& \frac{t_e}{2}\{[3(\vec{\sigma}_1 \cdot
\vec{k}^\prime)(\vec{\sigma}_2 \cdot \vec{k}^\prime) -
(\vec{\sigma}_1\cdot\vec{\sigma}_2){k^\prime}^2]\delta(\vec{r})
\notag\\
&+& \delta(\vec{r})[3(\vec{\sigma}_1 \cdot \vec{k})(\vec{\sigma}_2
\cdot \vec{k}) - (\vec{\sigma}_1\cdot\vec{\sigma}_2){k}^2]\}
\notag\\
&+&t_o[3(\vec{\sigma}_1 \cdot \vec{k}^\prime) \delta(\vec{r})
(\vec{\sigma}_2 \cdot \vec{k}) - (\vec{\sigma}_1\cdot\vec{\sigma}_2)
\vec{k}^\prime \cdot
\delta(\vec{r})\vec{k}]\notag\\
\end{eqnarray}
is incorporated to the full Skyrme transport model calculation,
where $\vec{r}=\vec{r}_1-\vec{r}_2$ is the relative coordinate,
$\vec{k}$ and $\vec{k}^\prime$ are the relative momentum operator
and its complex conjugate, respectively, and $t_e$ and $t_o$ are the
triplet-even and triplet-odd strength parameter. The energy density
function derived from the above tensor force can be found in
Refs.~\cite{Les07,Ben09}, where the corresponding terms (such as the
spin-current density $\vec{J}$) are non-negligible only when local
spin polarization is produced. The resulting spin up-down
differential transverse flow is shown in the right panel of
Fig.~\ref{full_tensor}. It is seen that the slope of $F_{ud}$ is not very
sensitive to the values of $t_e$ or $t_o$ unless extremely large
coupling constant is used. This feature is consistent with TDHF
study where the spin dynamics is dominated by the spin-orbit
coupling. However, one would expect that with spin-polarized beam or
target, the tensor force effect can be much enhanced.

\subsection{QMD model study}

In the QMD framework, the Wigner function of each nucleon is treated
as Gaussian wave packet in both coordinate and momentum
space~\cite{Har89,Aic91}, and the two-nucleon interaction is related
to the effective two-body interaction and the overlap of their wave
functions. The equation of motion in the QMD model is given by the
semiclassical canonical equation, i.e.,
\begin{eqnarray}
\frac{d\vec{r}}{dt} &=& \nabla_p H, \notag\\
\frac{d\vec{p}}{dt} &=& -\nabla_r H,
\end{eqnarray}
where $\vec{r}$ and $\vec{p}$ are respectively the central
coordinate and momentum of the wave packet, and $H$ is the
Hamiltonian of the system including the kinetic and potential
energy.

In a recent study, the nuclear spin-orbit interaction was
incorporated to the ultra-relativistic QMD (UrQMD) model. The
potential energy contribution of the spin-orbit interaction is
expressed as~\cite{SIQMD}
\begin{equation}
U_{s} = \int u_{s} d^3r,
\end{equation}
where the spin-dependent potential $u_s$ consists of the time-even
and time-odd contribution written as
\begin{eqnarray}
u_s^{even} &=& -\frac{W_0}{2}(\rho\nabla\cdot\vec{J} +
\rho_n\nabla\cdot\vec{J}_n +\rho_p\nabla\cdot\vec{J}_p ), \\
u_s^{odd} &=& -\frac{W_0}{2} [\vec{s} \cdot (\nabla \times \vec{j})
\notag\\
&+& \vec{s}_n \cdot (\nabla \times \vec{j}_n) + \vec{s}_p \cdot
(\nabla \times \vec{j}_p)],
\end{eqnarray}
where $W_0$ represents the spin-orbit coupling strength, and $\rho$,
$\vec{s}$, $\vec{j}$, and $\vec{J}$ are the number, spin, current,
and spin-current densities, which can be calculated from local
Wigner function of the nucleon~\cite{SIQMD}.

\begin{figure}[ht]
\includegraphics[scale=0.4]{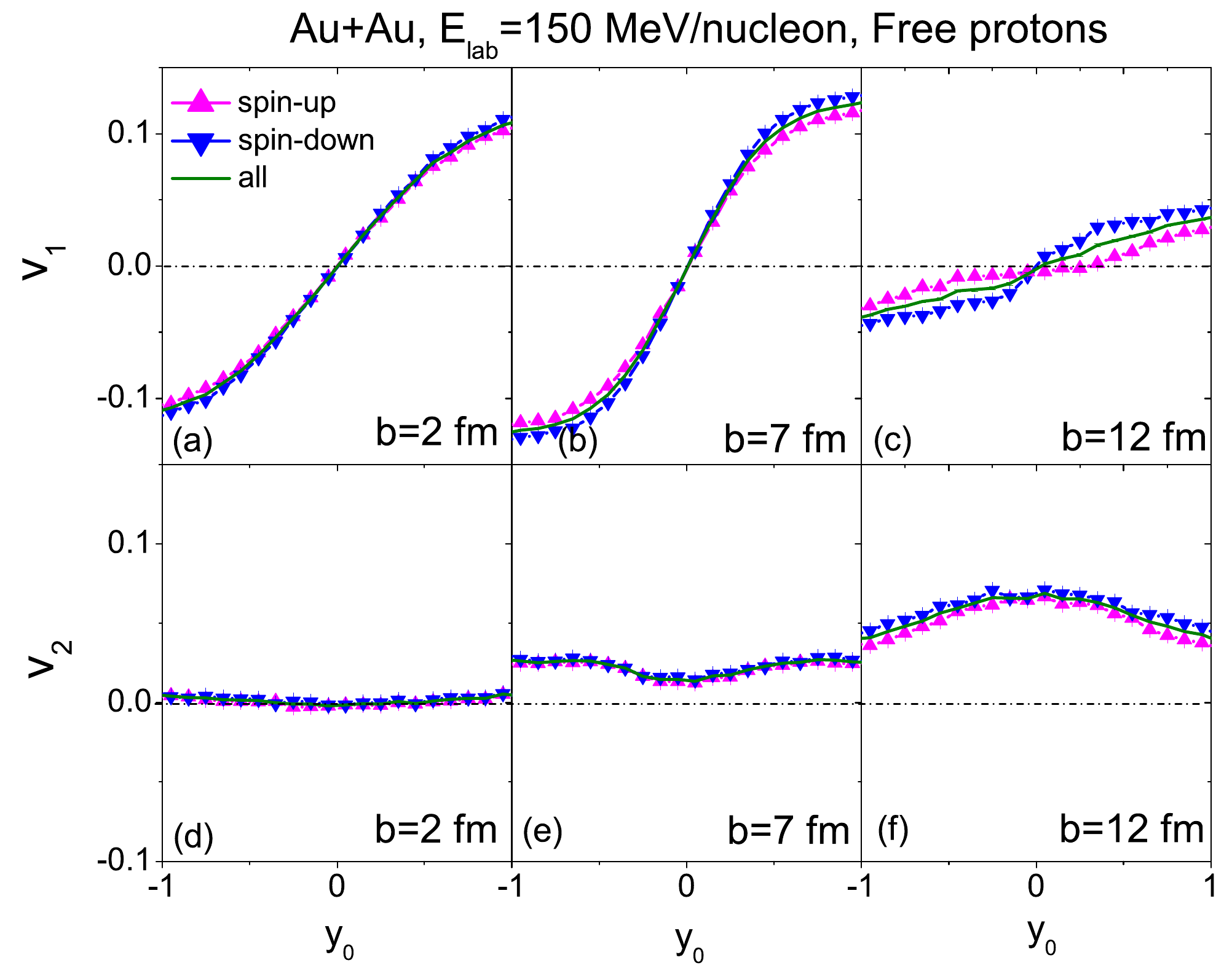}
\caption{(Color online) The directed flow (upper panels) and
elliptic flow (lower panels) for spin-up and spin-down protons in
non-central Au+Au collisions at the beam energy of $150$ MeV from
QMD calculation. Taken from Ref.~\cite{SIQMD}. }\label{qmd_v12}
\end{figure}

The spin dynamics was analyzed based on the above framework. Similar
spin splittings of the directed flow and the elliptic flow were
observed in non-central Au+Au collisions at the beam energy of $150$
MeV, as shown in Fig.~\ref{qmd_v12}. It was argued that the net
spin-dependent potential is attractive for spin-up protons and
repulsive for spin-down protons, leading to a larger directed flow
for spin-down protons than spin-up protons. The spin splitting of
$p_T$-integrated elliptic flow was found to be small and only
visible in peripheral collisions, and it was found that $v_2$ for
spin-down protons is slightly larger than that for spin-up ones.
Since the conventional initial direction of the target and
projectile is used as shown in Fig.~3 of Ref.~\cite{SIQMD}, the
spin-up (spin-down) nucleons correspond to the spin-down (spin-up)
ones in the SIBUU
study~\cite{SIBUU13,SIBUU14,SIBUU15,Xu14NPR,Xu14NT}. Although the
spin splitting of final collective flow is a robust phenomenon in
both models, further studies are needed to understand the relative
sign of the splitting.

\begin{figure}[ht]
\includegraphics[scale=0.3]{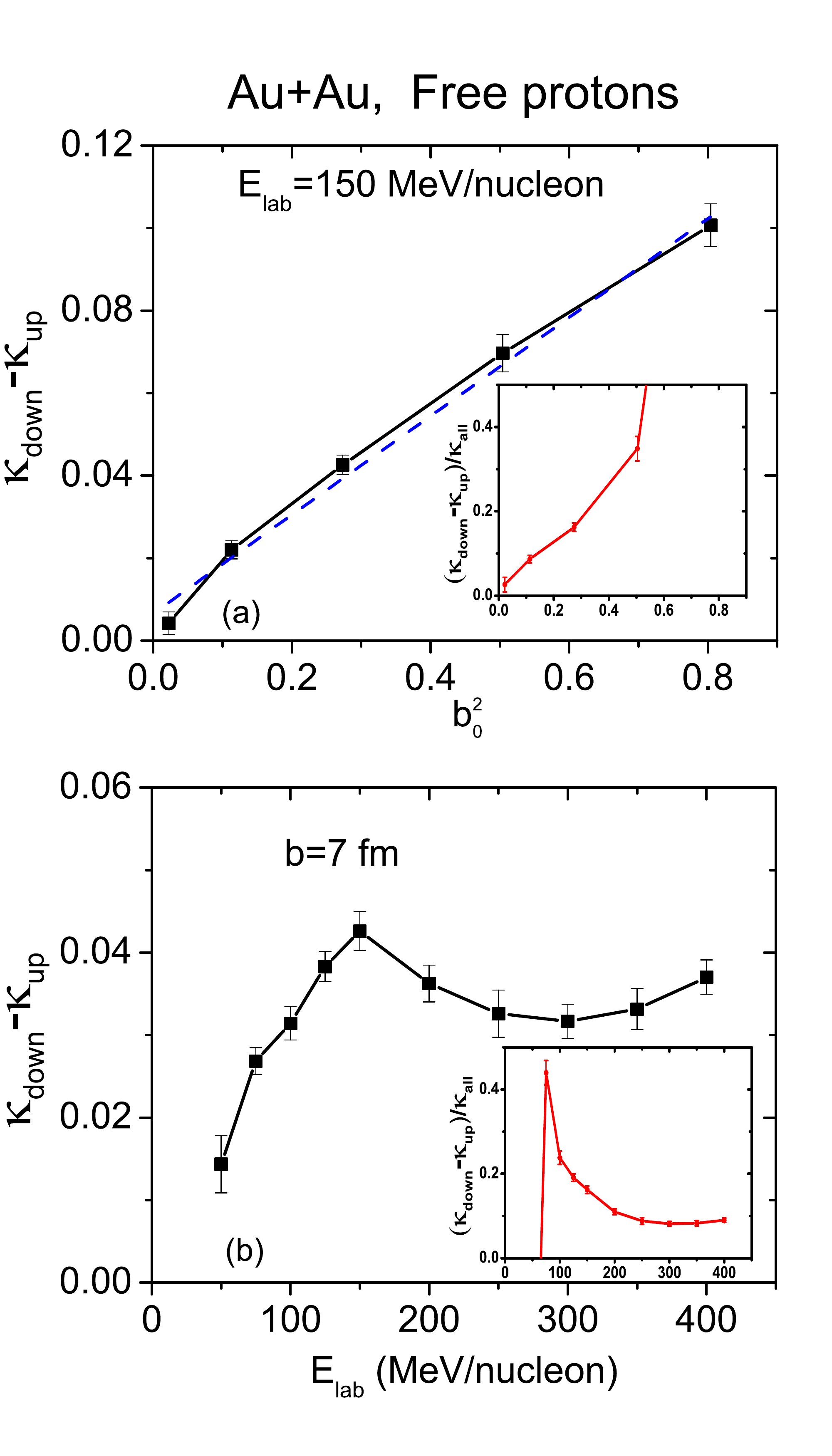}
\caption{(Color online) The difference of the slope of directed flow
between spin-up and spin-down protons as a function of impact
parameter (upper panel) and beam energy (lower panel). The dashed
line in the upper panel is a linear fit, while the inset in the
lower panel shows the relative difference. Taken from
Ref.~\cite{SIQMD}. }\label{qmd_slope}
\end{figure}

Analysis was further done on the beam energy dependence of the flow
splitting. Defining $\kappa_{up}$ and $\kappa_{down}$ as the slope
parameter of the directed flow of spin-up and spin-down protons, the
slope difference is shown to increase with increasing impact
parameter, as shown in the upper panel of Fig.~\ref{qmd_slope},
qualitatively consistent with SIBUU studies. In non-central Au+Au
collisions, it was found that the slope difference first increases
then decreases with increasing beam energy, and the maximum
difference appears at the beam energy of $150$ MeV, similar to the
finding in the SIBUU model where the maximum slope of the spin
up-down differential transverse flow appears at the beam energy of
about $100$ MeV.

It was further emphasized in Ref.~\cite{SIQMD} that the spin
averaged flow results do not change after including the spin-orbit
interaction, as a result of the cancellation of spin-up and
spin-down nucleons. In addition, the spin splitting of the flow
slope caused by the spin-orbit interaction is comparable to the
isospin splitting caused by the nuclear symmetry energy, especially
for neutrons. These findings are all consistent with the
observations in SIBUU
studies~\cite{SIBUU13,SIBUU14,SIBUU15,Xu14NPR,Xu14NT}.

\section{Experimental status}
\label{experiment}

Due to the difficulties of spin measurement in HIC experiments, the
main focus in the past is mainly on the spin-averaged observables,
so that the information of spin dynamics is neglected. Thanks to the
great efforts made by experimental nuclear physicists, the
measurements of the spin of free nucleons and light clusters now
become possible. Although the detailed experimental status will be
presented in another topic review of this issue, here we'd like to
briefly mention two related experiments that might be relevant in
analyzing the probes discussed above. One of them is the
spin-polarized beam which can be produced through pick-up or removal
reactions at Rikagaku Kenkyusho (RIKEN)~\cite{Asa90,Oku94},
Gesellschaft f\"{u}r Schwerionenforschung mbH (GSI)~\cite{Sch94},
the National Superconducting Cyclotron Laboratory
(NSCL)~\cite{Gro03}, and the Grand Acc\'{e}l\'{e}rateur National
d'Ions Lourds (GANIL)~\cite{Asa91,Bor02,Tur06}. It is expected that
the effects of spin dynamics with spin-polarized beam will be much
enhanced, providing a better system for extracting the information
of the spin-dependent nuclear force, especially the nuclear tensor
force. For the spin-excitation state of heavy clusters, the spin
polarization and alignment can be measured via the angular
distribution of its $\gamma$ or $\beta$ decay, see, e.g.,
Ref.~\cite{Ich12} for a review. Making use of the analyzing power of
a nucleus might be the most promising way of identifying the spin of
free nucleons or light clusters experimentally. The analyzing power
indicates the left-right scattering asymmetry of an incident
polarized nucleon on the target nucleus. The spin-dependent
scattering is a result of the interference of electromagnetic
interaction and hadronic force~\cite{But78}, and the spin flip is
observed between not only charged-charged scatterings but also
charged-neutral scatterings. It is noteworthy that at certain
energies and scattering angles the analyzing power can be as large
as 100$\%$~\cite{Zel11}. Experimental efforts are thus encouraged by
using the selected nucleus as a 'detector' whose analyzing power is
known in prior. In this way the spin of corresponding particles can
be measured and the probes discussed in the previous sections can be
analyzed.

\section{Summary}
\label{outlook}

In summary, we outlined the major physics motivations of investigating
the in-medium spin-dependent nuclear interactions, i.e., the spin-orbit
interaction and the nuclear tensor force, and summarized some recent efforts in
exploring the spin dynamics in low- and intermediate-energy
heavy-ion collisions. In particular, the studies on the strength,
the density, and the isospin dependence of the spin-orbit
interaction as well as the short-range correlation induced by the
tensor force are highlighted. In TDHF studies, it has been
found that the spin-orbit interaction can enhance the dissipation in
low-energy heavy-ion reactions and increase the fusion threshold.
Incorporating both the time-even and time-odd contribution of the
spin-orbit interaction can lead to nontrivial spin polarization,
while the tensor force slightly enhances the spin field compared to
the spin-orbit interaction. In the studies using HICs at intermediate energies, both the spin- and isospin-dependent BUU model and QMD model
predict the spin splitting of the nucleon collective flow, which
seems to be a robust phenomenon. In the BUU model studies, efforts
have been made in extracting the isospin dependence of the
spin-orbit coupling and disentangle its strength and density
dependence. Preliminary results on spin splitting of observables
related to the light clusters and those from full Skyrme calculation
with nuclear tensor force have also been discussed in the BUU model
studies.  We hope the findings summarized in this review will soon stimulate more
experimental work in this direction.

\begin{acknowledgments}
This work was supported by the Major
State Basic Research Development Program (973 Program) of China under Contract Nos. 2015CB856904 and
2014CB845401, the National Natural Science Foundation of China under
Grant Nos. 11320101004, 11475243, and 11421505, the "100-talent plan" of
Shanghai Institute of Applied Physics under Grant Nos. Y290061011 and Y526011011
from the Chinese Academy of Sciences, the "Shanghai Pujiang Program"
under Grant No. 13PJ1410600, the US National Science Foundation under Grant No. PHY-1068022, the U.S. Department of Energy Office of Science under Award Number DE-SC0013702,
and the CUSTIPEN (China-U.S. Theory Institute for Physics with Exotic Nuclei) under the US Department of Energy Grant No. DE-FG02-13ER42025.

\end{acknowledgments}

\end{document}